\documentstyle[preprint,aps]{revtex}

\begin{document}

\title{Testing quantum superpositions of the gravitational field with
Bose-Einstein condensates}

\author{Netanel H. Lindner and Asher Peres}
\address{Department of Physics, Technion---Israel Institute of
Technology, 32000 Haifa, Israel}

\bigskip

\def\6{\langle}
\def\9{\rangle}
\def \beq{\begin{equation}}
\def\eeq{\end{equation}}
\def\bea{\begin{eqnarray}}
\def\eea{\end{eqnarray}}
\def\half{\mbox{$1\over2$}}
\def\halfb{\mbox{$\beta\over2$}}
\def\halft{\mbox{$\theta\over2$}}
\def\qtr{\mbox{$1\over4$}}
\def\ses{\mbox{$1\over16$}}
\def\br{\textbf{r}}
\maketitle
\title{Testing Quantum Superposition of the Gravitational Field with Bose-Einstein Condensates}

\begin{abstract}
We consider the gravity field of a Bose-Einstein condensate in a
quantum superposition. The gravity field then is also in a quantum
superposition which is in principle observable. Hence we have
``quantum gravity'' far away from the so-called Planck scale.
\end{abstract}
\pacs{PACS numbers: 03.65.-w}
\maketitle

The existence of macroscopically distinguishable superpositions in
Bose-Einstein condensates (BEC) have been discussed by several
authors \cite{cirac,ruos,dalvit}. Such a superposition may be
achieved as the ground state of two interacting BECs in a double
well potential \cite{cirac}, or by a continuous quantum
measurement on a condensate trapped in such a
potential\cite{ruos}.


We now consider the gravitational field of a BEC (Bose-Einstein
condensate) of total mass M in such a double well potential and we
scatter a beam of particles of mass $m$ in the middle of the
double potential well. (We assume that the potential that affects
the BEC does not effect these particles, or that we know how to
correct for its effect in our calculations).

In order to avoid decoherence, the density matrix of the scattered
particles must be factorized in the form: $\rho=\rho_{\rm
int}\otimes\rho_{\rm CM}$ where $\rho_{\rm int}$ stands for the
density matrix of the internal degrees of freedom, and $\rho_{\rm
CM}$ for the density matrix of the external degrees of freedom.

Let us treat classically the interaction of the center of mass of
the scattered particles with the gravitational field of the BEC.
Suppose that we know for certain that the BEC condensate is
localized within the left well or the right well (this may happen,
for example, in the case of an infinite barrier in the double
potential well). The scattered particles will be deflected from
their course due to the gravitational attraction between them and
the BEC. Here we also assume that all other interactions between
the scattered particles and the BEC either do not exist or are
negligible with respect to the gravitational interaction (which
for itself is very weak). We shall set up our axis system so that
the double potential well lies on the x axis (the maximum of the
potential barrier is at $x=0$), and the scattered particiles
initial trajectory is along the y axis, with $x=0$, so that their
inital momentum is $p_0=(0,p_y,0)$. The distances between the two
minima in the potential will be $a$, and the particles
trajectories are such that without the gravitational interaction
they would pass at a distance $a$ within each minima. In the
crudest, ``undergraduate" approximation, the scattered particles,
will be under the influence of a gravitational force $GmM/a^2$
during a time interval of $a/v_y=a m p_y$. The scattered
particles, initially with $p_x=0$ will have, after the
interaction, a non zero $x$ component for their momentum $\pm
GMm^2/ap_y$, where $+$ is for the case in which the BEC is in the
right well, and $-$ if it was on the left one. The deflection
angle, in each case will be small and can be written as $\theta
\approx\pm
p_x/p_y=GMm^2/ap_y^2=GM/av_y^2$.

Now assume that the potential barrier between the two wells is
finite and the BEC condensate is in a symmetric state
$\psi_S=\psi_L+\psi_R$ where $\psi_R$ and $\psi_L$ are functions
that are localized in the right and left wells respectively. What
is the result of scattering  the particles from the gravitational
field of this state? After the scattering, the state of the center
of mass of the particles will be entangled with he state of the
BEC and in a  superposition of states, one being a result of a
deflection from a BEC in the right well, and the other being the
result of deflection from a BEC in the left well. The state of the
system can be written in the following manner:
\beq
\varphi(\br)=A_1\, \psi_L\otimes \exp\left(i/\hbar(p_x,p_y,0)\cdot \br\right) +
 A_2\, \psi_R \otimes\exp\left(i\hbar(-p_x,p_y,0)\cdot \br\right)
\eeq
Here it is essential that the initial momentum spread $\Delta p$
of the BEC be much larger than the momentum kick due to the
interaction.

We do not record the state of the BEC during the experiment, or in
other words we are tracing over the BEC states. If $\psi_L$ and
$\psi_R$ were orthogonal, tracing over the BEC state would prevent
us from seeing any interference fringes. The situation is
different if they are not orthogonal. Let $\xi=\6\psi_L|\psi_R\9.$
The density matrix of the center of mass of the scattered
particles after tracing over the BEC states is
\bea
 \rho(\br,\br') &=& |A_1|^2\exp [i/\hbar(  p_x,p_y,0)\cdot({\bf r-r'})] \nonumber \\
 &+& |A_2|^2\exp [i/\hbar(-  p_x,p_y,0)\cdot({\bf r-r'})] \nonumber\\
 &+& 2{\rm Re}\{A_1A_2^*\xi \exp i/\hbar[( _x,p_y,0)\cdot{\bf
 r}]\nonumber\\
& +&(  p_x,-p_y,0)\cdot {\bf r'}]\}.
\eea
Assuming $A_1=A_2$, it is easy to see that
$\rho\left(\br,\br\right)$ will be maximal when
$\cos(p_xx/\hbar)=1$, or $p_xx/\hbar = 2\pi m$, so the distance
between fringes will be
\beq
\Delta x = 2\pi \hbar /p_x = (h a v_y)/(G M m)
\eeq

Let us  try to see the parameters needed in such an experiment in
order for it to be feasible. As we shall soon see, the distance
between fringes tends to be very large, so let us choose
parameters as to make it as small as possible (while difficult, it
does not violate any law of physics). A BEC usually contains up to
$10^7$ Rb atoms, which give a mass of about $10^{-18}$ Kg. For the
scattered particles, let us take big particles (grains) with a
mass of about a nanogram, or $10^{-12}$ Kg. The constants $h$ and
$G$ have values

\begin{displaymath}
h\approx 6\times10^{-34} {\rm Joule\;sec}
\end{displaymath}
\begin{displaymath}
G \approx 6\times10^{-11} {\rm \;Kg^{-1}\; m^3\; sec^{-2}}
\end{displaymath}
We need to get the product $a v_y$ as small as possible. The
distance between the minima in the double potential well cannot be
much smaller than about 1 micron, or $10^{-6}$ meters. The
question now is, how slow can the incoming particles be and still
scatter coherently only from the gravitational field of the BEC?
Introducing all the above numbers, we have
\beq
\Delta x = 10\;v_y \;{\rm[meters]}
\eeq
So in order to achieve a distance between fringes that is on the
order, say, of cm, so that the experiment will be feasible we need
the incoming particles speed to be of the order $10^{-3}$
meters/sec. These are slow speeds, but nevertheless, there is no
physical law that prevents them. We can also try to take heavier
particles for the scattered particles. We can write:
\beq
\Delta x = 10^{-11}(v_y/m)\;{\rm[meters]}.
\eeq
The ratio $v_y/m$ will have to be of the orders of $10^9$
meters/(kg sec).


\begin{thebibliography}{99}
\bibitem{cirac} J. I. Cirac, M. Lewenstein, K. M$\o$lmer and P.
Zoller, Phys. Rev. A. {\bf 57}, 1208 (1998).

\bibitem{ruos} J. Ruostekoski, Lecture Notes in Physics: Directions in Quantum Optics, (Springer, Berlin, 2001),
 cond-mat/0005469.
\bibitem{dalvit}  D. A. R. Dalvit, J. Dziarmaga, and W. H. Zurek, Phys. Rev. A {\bf 62},
013607 (2000).
\end{thebibliography}
\end{document}